\PassOptionsToPackage{hyphens}{url}

\documentclass[letterpaper,twocolumn,10pt]{article}

\makeatletter
\newcommand{\DeclareLatinAbbrev}[2]{%
  \DeclareRobustCommand{#1}{%
    \@ifnextchar{.}{\textit{#2}}{%
      \@ifnextchar{,}{\textit{#2.}}{%
        \@ifnextchar{!}{\textit{#2.}}{%
          \@ifnextchar{?}{\textit{#2.}}{%
            \@ifnextchar{)}{\textit{#2.}}{%
              {\textit{#2.,\ }}}}}}}}%
}
\makeatother
\DeclareLatinAbbrev{\eg}{e.g}
\DeclareLatinAbbrev{\Eg}{E.g}
\DeclareLatinAbbrev{\ie}{i.e}
\DeclareLatinAbbrev{\Ie}{I.e}
\DeclareLatinAbbrev{\etc}{etc}
\DeclareLatinAbbrev{\etal}{et~al}

\def\first {$(i)$\xspace}
\def\second{$(ii)$\xspace}
\def\third {$(iii)$\xspace}

\newcommand{\smartparagraph}[1]{\vspace{.01in}\noindent\textbf{#1}}

\DeclareRobustCommand{\circled}[1]{\tikz[baseline={([yshift=-1pt]char.base)}]{
            \node[shape=circle,draw=black,dash pattern=on 0.4pt off 1pt,fill=white,inner sep=1pt, minimum size=12pt,scale=0.9] (char) {#1};}}

\DeclareRobustCommand{\circleddark}[1]{\tikz[baseline=(char.base)]{
            \node[shape=circle,fill=black,inner sep=0pt, minimum size=12pt,scale=0.9] (char) {\textcolor{white}{#1}};}}

\newcommand{\systemname}{\textsc{Spillway}\xspace}
\newcommand{\ATOA}{AllToAll\xspace}
\newcommand{\AR}{AllReduce\xspace}

\makeatletter
\let\oldthanks\thanks
\renewcommand{\thanks}[1]{%
  {\hypersetup{hidelinks}\oldthanks{#1}}%
}
\makeatother
\newcommand{\affmark}[1]{$^{#1}$}

\usepackage{usenix2019_v3}

\usepackage{tikz}
\usepackage{amsmath}
\usepackage{natbib}

\usepackage[utf8]{inputenc} 
\usepackage[T1]{fontenc}    
\usepackage{url}            
\usepackage{booktabs}       
\usepackage{amsfonts}       
\usepackage{nicefrac}       
\usepackage{microtype}      
\usepackage{xcolor}         
\usepackage{cite}
\usepackage{graphicx}
\usepackage{float}
\usepackage{xspace}
\usepackage{tabularx}
\usepackage{longtable} 
\usepackage{colortbl}
\usepackage{siunitx}
\usepackage{multirow}
\usepackage{makecell}
\usepackage{dcolumn}
\usepackage{pifont}
\usepackage{placeins}
\usepackage{tikz}
\usepackage{xspace}
\usepackage{cleveref}
\usepackage{subfig}
\usepackage{enumitem}
\usepackage{tcolorbox}
\usepackage{lastpage}
\usepackage{todonotes}

\title{Avoiding Cross-Datacenter Collective Congestion via Disaggregated Buffering}

\author{
Mariano Scazzariello\affmark{1},  
Noga H. Rotman\affmark{2},  
Dima Gavrilenko\affmark{3},
Sajy Khashab\affmark{4},\\
Alexander Shpiner\affmark{4},
Matty Kadosh\affmark{4},
Marco Chiesa\affmark{5},
Dejan Kostic\affmark{5},
Mark Silberstein\affmark{3}
\\[0.5em]
\parbox{\linewidth}{\centering\footnotesize
\affmark{1}RISE Research Institutes of Sweden, 
\affmark{2}University College London, 
\affmark{3}Technion - Israel Institute of Technology, \\
\affmark{4}NVIDIA, 
\affmark{5}KTH Royal Institute of Technology
}
}

\begin{document}

\maketitle

\begin{abstract}
LLM training at the scale of tens of thousands of GPUs now spans multiple datacenters (DC), making cross-DC collectives over long-haul links unavoidable. A critical and overlooked bottleneck arises when these collectives collide with intra-DC traffic at the destination -- a common pattern in real workloads. The 
multi-millisecond congestion control loop is too slow to react, triggering severe packet loss and congestion collapse.

We present \systemname, a transparent in-network mechanism that buffers dropped packets in switch-disaggregated buffers in a destination data center and drains them once congestion subsides. Through large-scale end-to-end simulations and a hardware prototype, we show that \systemname eliminates performance degradation from collective collisions, reducing iteration time by up to 14\,\%, without changes to end hosts or training frameworks.

\end{abstract}

\section{Introduction}\label{sec:intro}


Cross-DC training is necessary to sustain the scale of modern AI systems, driven in large part by the rapidly growing energy demands of training.
For example, recently it was shown that single-site DC designs become infeasible beyond a certain scale: no location within a 100\,km radius offers sufficient power and cooling to support trillion-parameter models such as Claude Mythos~\cite{claude_mythos}, forcing operators to split their infrastructure across multiple DCs~\cite{gherghescu2024ve}. 
This shift is happening in practice: xAI is expanding its Colossus 2 cluster across multiple nearby sites~\cite{semianalysis_colossus2}, while frontier labs such as Google and OpenAI already trained models on geo-distributed clusters~\cite{gpt45pretrain,geminiteam2025geminifamilyhighlycapable,semianalysis_multidc}.

\begin{figure}[t]
    \vspace{-.2in}
    \centering
    \includegraphics[width=\linewidth]{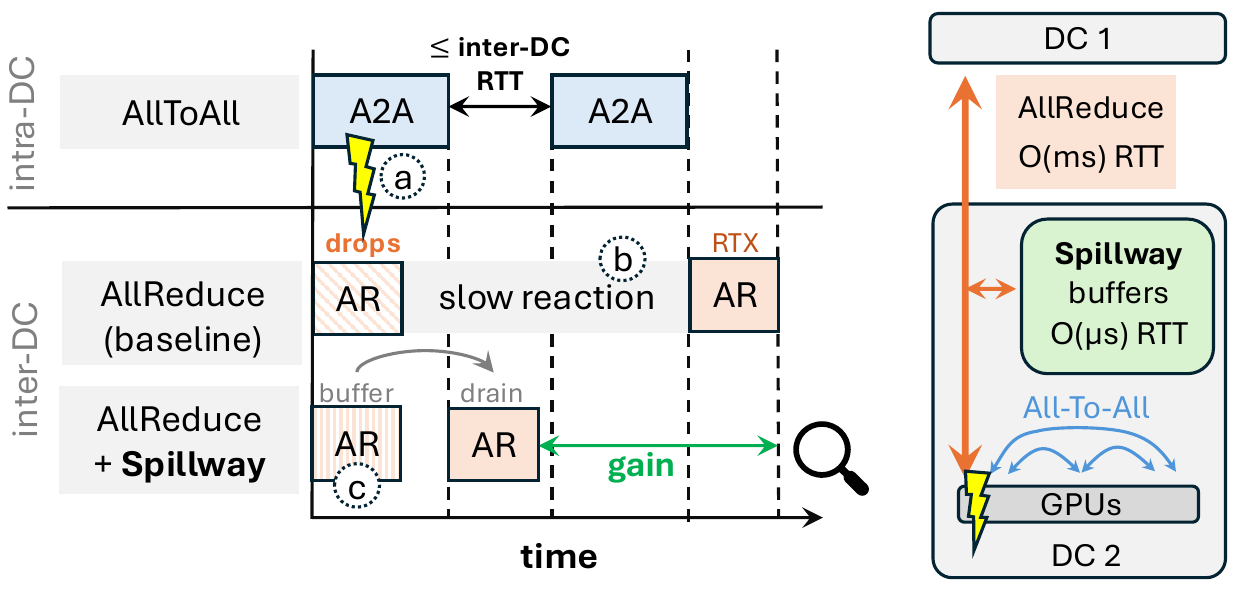}
    \vspace{-.2in}
    \caption{Cross-DC \AR traffic collides with bursty intra-DC \ATOA at the destination, causing buffer overflow, packet drops, and slow RTT-scale recovery. \systemname buffers dropped packets and drains them after the burst, avoiding retransmissions and improving completion time.}
    \label{fig:netsponge-collectives-collide}
\end{figure}

\smartparagraph{When collectives collide.} Cross-DC latency is orders of magnitude larger than intra-DC latency. Growing roughly 5 milliseconds every 1,000~km, it reaches  tens of milliseconds for  cross-DC links versus microseconds within a rack~\cite{si2026collectivecommunication100kgpus,nvidia_nemo_multidc}. When training an ML model, common production deployments place data parallel replicas across DCs~\cite{nvidia_nemo_multidc,gherghescu2024ve, si2026collectivecommunication100kgpus}, causing gradient aggregation to traverse long-haul links. These cross-DC transfers inevitably overlap with intra-DC collectives, such as \ATOA, that execute concurrently within each site~\cite{gao2025flowmoescalablepipelinescheduling,288705}.
At the receiver, this overlap creates a fundamental timescale mismatch. Intra-DC collectives generate short, high-rate bursts that monopolize the destination port, while cross-DC traffic arrives as long-haul flows with slow feedback. Ideally, remote packets would be scheduled into the small idle gaps between local bursts. In practice, these gaps are \textit{insufficient for a cross-DC RTT-scale feedback loop to react}.
As a result, cross-DC traffic is injected into the destination DC without coordination with local bursts. When it collides with a local burst (see \circled{a} in Fig.~\ref{fig:netsponge-collectives-collide}), packets accumulate at the destination leaf switch and quickly overflow buffers, leading to drops, often within sub-millisecond timescales.

\smartparagraph{RDMA for cross-DC traffic: amplifying the mismatch.} The problem is further exacerbated by the industry’s adoption of RDMA 
for ML training~\cite{si2026collectivecommunication100kgpus,gangidi2024rdma,qian2024alibaba}. RDMA was designed for tightly coupled, low-latency environments, where fast feedback and shallow buffers enable efficient loss recovery. Across DCs, these assumptions break down: a single packet drop can 
turn a short-lived collision into a multi-millisecond stall, leading to a \textit{slow reaction} (see \circled{b} in Fig.~\ref{fig:netsponge-collectives-collide}).

In response, operators have moved away from traditional RDMA congestion control (CC). Meta reports disabling DCQCN in its 100K-GPU RoCE fabric for Llama~4 training, replacing it with library-level receiver-driven flow control~\cite{si2026collectivecommunication100kgpus}; Alibaba similarly adopts a custom CC design~\cite{qian2024alibaba}. A parallel line of work improves loss recovery through mechanisms such as selective retransmission and multipath transport~\cite{li2025dcp,mittal2018revisiting,mprdma}. These approaches mitigate persistent oversubscription and better match large bandwidth-delay products, but they remain fundamentally reactive. Operating on millisecond timescales, however, they still cannot prevent the microsecond-scale collisions and packet loss that occur in multi-DC training, thus unable to mitigate costly retransmissions  that further inflate completion time.

\smartparagraph{Deep buffers absorb microbursts, at a price.}
The only mechanism that directly absorbs microsecond-scale bursts is deep buffering, which Meta deploys alongside its flow control scheme~\cite{si2026collectivecommunication100kgpus}. While effective, deep buffers are costly and rigid: HBM-backed ASICs are significantly more expensive per port, consume substantial power, and must be provisioned for worst-case collisions that occur only intermittently. Because buffer capacity is fixed at deployment time, a configuration that is sufficient today can quickly become inadequate as collectives, models and cross-DC distances scale, forcing overprovisioning or costly upgrades. Burst absorption is unavoidable; the key question is whether this capacity should reside inside every switch or can instead be \emph{disaggregated} into a flexible pool shared across the DC.




\begin{figure}[b!]
    \vspace{-.2in}
    \centering
    \includegraphics[width=.8\linewidth]{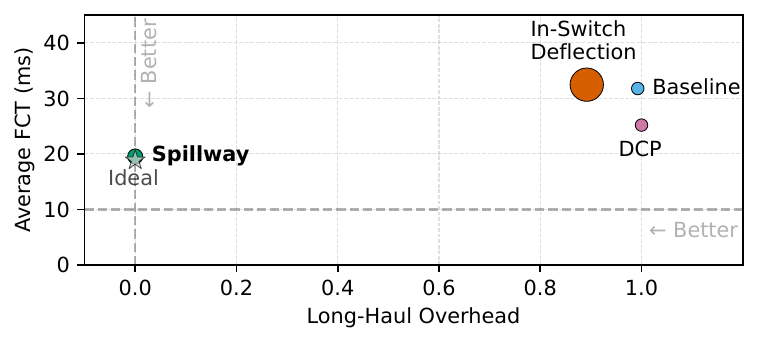}
    \vspace{-.15in}
    \caption{Different approaches with 16 concurrent remote flows and a local \ATOA. Points show average FCT vs.\ long-haul overhead,  circle size proportional to deflection overhead.}
    \label{fig:efficiency_motiv}
\end{figure}

\smartparagraph{\systemname: disaggregated buffer for cross-DC traffic.} We propose a simple principle: once a packet has traversed the inter-DC link, it should not be dropped at the destination boundary. Dropping it shifts recovery to the sender, which must retransmit the packet and \emph{time its arrival} to slip into an idle gap between local bursts, an intractable timing problem given hundreds of microsecond-scale gaps and tens of millisecond-scale cross-DC RTTs. Instead, the destination DC should \emph{absorb} these collisions locally and release packets once the port drains (\circled{c} in Fig.~\ref{fig:netsponge-collectives-collide}). Because both buffering and control occur inside the DC, this loop operates at microsecond timescales, enabling timely reinjection of packets between local bursts, which no cross-DC controller can achieve.

\systemname realizes this principle by performing burst absorption into a \emph{disaggregated buffer} at the DC boundary. When a switch would drop a remote packet (see right side of Fig.~\ref{fig:netsponge-collectives-collide}), the switch instead deflects it to a special location(s), which we call a \emph{spillway node}. Such a node can be a commodity SmartNIC connected to a switch or even a server whose memory acts as an extension of the switch buffer. Packets are temporarily held at the destination DC and later reinjected at the right time, avoiding slow cross-DC retransmissions, transparently for applications.

\smartparagraph{Comparison to existing approaches.}
Unlike prior deflection schemes~\cite{zarifis2014dibs,abdous2023practical,abdous2021burst}, which deflect packets towards other switches and spread congestion across the network, \systemname \emph{stops} packets at a safe location and releases them only when they can make forward progress, avoiding both cross-DC retransmissions and in-network churn.

Fig.~\ref{fig:efficiency_motiv} illustrates the performance relationship between different designs. We simulate a dual-DC setup with 32 GPUs per DC, where a high\nobreakdash-priority local \ATOA collective collides with 16 remote flows, emulating cross-DC hierarchical \AR traffic, under realistic training workloads conditions (Sec.~\ref{sec:eval} for more details). The baseline 
ends up retransmitting all the remote flow twice. Deflection-based solutions incur substantial overheads, either due to retransmissions or repeated deflections, making them ill-suited for this setting. \systemname occupies the ideal point in the design space, with minmimum deflections, no retransmission overheads and optimal completion time. 

\smartparagraph{Challenges.} While conceptually simple, realizing \systemname in practice introduces several challenges. 
\first~the deflection of \textit{entire} packets must be below the transport layer to avoid application changes; \second~without a proper CC mechanism, the spillway nodes resending the traffic may reintroduce incast at the destination port; conversely, deflected packets may stall indefinitely in the spillway, as there is no mechanism to trigger their timely retrieval; \third~there are multiple spillway nodes around the cluster among which the deflected packets must be load-balanced. 


\systemname tackles these challenges by carefully coordinating \textit{where} packets are buffered, \textit{when} they are reintroduced into the network, and \textit{how} deflected traffic is distributed across spillway nodes. It places external buffers at the DC boundary (\ie exit switches) to avoid interfering with ongoing traffic, uses lightweight probing and timing mechanisms to safely re-inject packets once congestion clears, and balances deflected traffic across multiple spillways to avoid creating new hotspots. Together, these mechanisms ensure that packets are buffered without loss, reinjected without disruption, and distributed without introducing new bottlenecks. 

We extensively evaluate \systemname through simulations and a hardware prototype, demonstrating that it eliminates retransmissions and stabilizes cross-DC performance. In realistic training workloads, it reduces microbatch time by up to 14\,\% and overall iteration time by up to $\sim$5\,\%. On the real prototype, it achieves up to 40\,\% FCT reduction.

\smartparagraph{General benefits of disaggregated buffering.}
More broadly, we view disaggregated buffering as a new design point that decouples buffer capacity from fixed switch resources. Instead of statically provisioning per-port buffers, operators can allocate capacity where and when it is needed as workloads evolve (\eg new collectives, traffic patterns, or topologies). This flexibility avoids both overprovisioning switch memory and costly hardware upgrades when demands increase. While we demonstrate these benefits in the context of cross-DC collectives, we believe this abstraction applies more broadly to dynamic DC workloads; a deeper exploration of these opportunities is left for future work.

\smartparagraph{Contributions.} Our contributions are as follows:
\begin{itemize}[leftmargin=*,noitemsep,topsep=0pt,parsep=0pt,partopsep=0pt]
    \item 
    We show that cross-DC LLM training suffers from a fundamental timescale mismatch: the long inter-DC RTT is far larger than the idle gaps between local collective bursts, so cross-DC flows cannot be scheduled into those gaps and instead collide with them, causing packet drops and slow long-haul recovery.

    \item 
    We design \systemname, which is more than just an extended switch buffer. Beyond holding packets at spillway nodes, it implements the control logic needed to reinject them safely: deciding when to drain, probing the destination path, and load-balancing across spillways. All of this operates transparently below the transport layer, with no changes to congestion control or collective libraries.

    \item 
    We evaluate \systemname through extensive end-to-end training simulations and a real hardware prototype built on NVIDIA Spectrum-4 switches and BlueField-3 DPUs. In simulations, it reduces microbatch time by up to 14\,\% and iteration time by up to $\sim$5\,\%. On the hardware testbed, it achieves up to 40\,\% FCT reduction.
\end{itemize}

\section{Background}\label{sec:background}


\vspace{-.05in}
\smartparagraph{Cross-DC training topologies.} Multi-site AI clusters 
interconnect multiple DCs via high-capacity DCI networks, often backed by dedicated backbone infrastructure and hierarchical aggregation layers~\cite{microsoft_backbone,semianalysis_multidc,meta_prometheus}. These deployments can span hundreds to thousands of kilometers to accommodate power and cooling constraints~\cite{semianalysis_multidc}, resulting in inter-site latencies on the order of tens of milliseconds, several orders of magnitude higher than intra-DC latencies. 
Unlike the non-blocking fabrics commonly assumed within a single DC~\cite{gangidi2024rdma,qian2024alibaba,nvidia_dgx_b200_ra,azure_gb300_blog}, cross-site networks operate under controlled oversubscription; \eg Meta reports a $\sim$4.5:1 oversubscription ratio in large-scale aggregation tiers~\cite{meta_prometheus}. At the same time, the absolute bandwidth between training clusters may reach Petabytes/second to achieve higher training efficiency~\cite{meta_prometheus}.

\smartparagraph{Intra-training congestion.} In LLM training  multiple collective operations are pipelined to maximize accelerator utilization~\cite{288705,10.1145/3458817.3476209}. While effective, this overlap introduces interference across collectives. For example, when MoE \ATOA overlaps with concurrent \AR, \ATOA can slow down by up to $4\times$; prioritizing \ATOA mitigates this effect and yields up to $1.73\times$ speedup~\cite{288705}.
This challenge is amplified in cross-DC training, where additional inter-site collectives increase both traffic volume and the likelihood of collisions. In practice, cross-DC communication is treated as lower priority to protect intra-DC collectives on the critical path. However, enforcing this separation is difficult: sustaining lower-priority long-haul traffic requires buffering proportional to the cross-DC BDP, which by far exceeds commodity switch capacity. 
As a result, packet drops and retransmissions over cross-DC links directly extend iteration time. This creates a fundamental tension: training demands prioritization across overlapping collectives, but the network cannot reliably enforce it with conventional buffering alone.

\smartparagraph{Multi-site-aware collectives.} Collective communication in multi-DC training relies on specialized designs, e.g., Hierarchical AllReduce, \emph{HAR}, that partition ranks by site and structure operations in multiple stages (\eg local ReduceScatter/AllGather followed by a smaller cross-site phase) to limit long-haul traffic~\cite{nvidia_nemo_multidc}. 
Recent work shows that collective performance is highly \mbox{topology-dependent} even within a single cluster~\cite{10.1145/3651890.3672249,285084,10.1145/3718958.3750499}, making explicit site-awareness essential at multi-DC scale. As a result, such designs are a key building block for scaling training throughput under realistic deployment constraints.

\section{Motivation}\label{sec:motivations}

\vspace{-.05in}

\smartparagraph{Cross-DC collisions inflate FCTs, delaying iterations.} Collisions occur when low-priority cross-DC collectives arrive at a destination concurrently executing high-priority intra-DC collectives. Because these operations overlap and contend for the same switch resources, long-haul traffic accumulates behind bursty local communication, leading to buffer overflow and packet drops. Unlike intra-DC congestion, these events are costly: large RTTs make recovery slow, so even a few losses significantly inflate FCTs and delay iterations.

To illustrate this effect, we simulate a dual-DC setup in ns-3 with 32 GPUs per site, 5\,ms one-way DCI latency, and $2\times400$\,Gbps cross-DC links per exit switch (see Sec.~\ref{ss:eval-sim} for details). We consider a 250\,MB long-haul HAR flow colliding with a 4\,GB local \ATOA at the destination DC. Both are derived from an MLSynth~\cite{10.1145/3748273.3749211} trace of a 24\,B-parameter sparse MoE model with 0.6\,B active parameters. The \ATOA runs across eight GPUs connected to the same leaf switch, and the RTO is set to 16.8\,ms, close to the long-haul RTT~\cite{ibta:spec}.

\begin{figure}[b!]
    \vspace{-.2in}
    \centering
    \includegraphics[width=.85\linewidth]{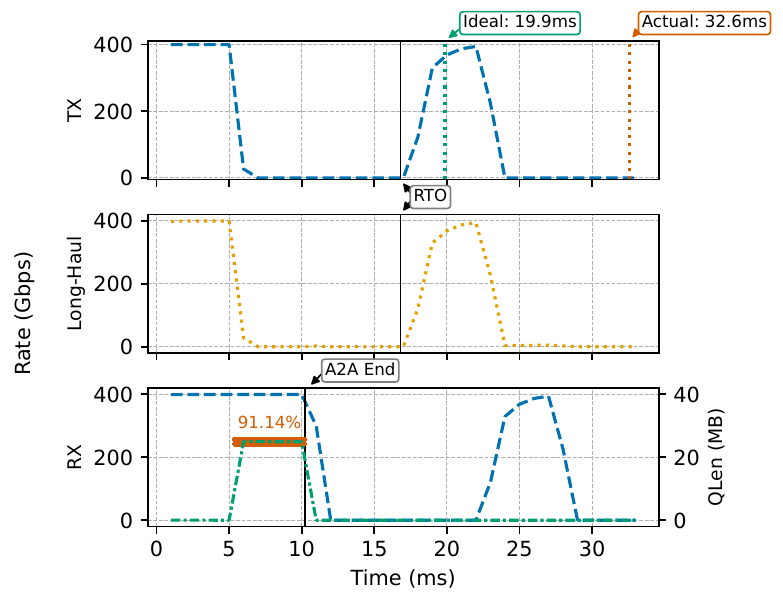}
    \vspace{-.15in}
    \caption{Impact of long-haul loss under cross-DC traffic (5\,ms one-way latency). A long-haul flow overlapping with a local \ATOA experiences buffer buildup, severe packet loss ($\sim$91\,\%), and delayed recovery via retransmission timeouts, increasing FCT from 19.9\,ms to 32.5\,ms (1.64$\times$).}
    \label{fig:ar_motiv}
\end{figure}

Fig.~\ref{fig:ar_motiv} shows that the long-haul flow (dashed) arrives while the \ATOA is in progress and is unable to make progress. Packets accumulate in the switch buffer (dash-dotted), which quickly exceeds 20\,MB before overflowing, resulting in $\sim$91\% loss. Although the \ATOA completes around 10\,ms, recovery is delayed by the RTT-scale feedback loop: about 90\% of the flow is retransmitted, inflating FCT to 32.5\,ms, a 1.64$\times$ slowdown compared to the ideal 19.8\,ms.

\smartparagraph{Hierarchical collectives amplify stragglers.} The previous example considers a single cross-DC flow. In practice, hierarchical collectives such as Hierarchical \AR~\cite{nvidia_nemo_multidc} (HAR) generate many concurrent cross-DC transfers to the same destination, increasing the likelihood of loss during collisions. Because HAR is synchronization-bound, iteration time is dictated by the slowest flow: a single straggler delayed by retransmissions stalls the entire collective. As the number of cross-DC flows grows, so does the probability of such stragglers, amplifying the impact of collisions.

\smartparagraph{Packet loss is prohibitively expensive at cross-DC scale.} Extending RDMA across DCs breaks its core assumptions. Long-haul links exhibit large BDPs and millisecond-scale RTTs~\cite{mdpi_rdma_survey}, making lossless operation impractical due to prohibitive buffer requirements. For example, following vendor guidelines~\cite{cisco-pfc}, a 400\,Gbps link over 5\,ms one-way latency requires $\approx$500\,MB of buffering \textit{per lossless queue}, exceeding commodity switch capacity~\cite{switch-buffers}.
As a result, cross-DC deployments operate in lossy mode and rely on end-host recovery (timeouts or selective retransmissions)~\cite{mdpi_rdma_survey,li2025dcp,mprdma,mittal2018revisiting}. While recent designs improve recovery latency and tolerate out-of-order delivery~\cite{mittal2018revisiting,mprdma,li2025dcp,aws_semianalysis}, they remain reactive: losses are repaired via retransmissions. In our setting, where collisions induce bursty losses, this requires retransmitting substantial data over high-RTT paths, making recovery fundamentally RTT-bound. Consequently, \textit{avoiding} drops, rather than \textit{recovering} from them, becomes the primary objective.

Two alternatives can achieve loss avoidance with existing mechanisms: preventing collisions at the application layer through \textit{scheduling}, or reacting to them in the network through \textit{deflection} among switches. We show why both fall short.

\smartparagraph{Scheduling cross-DC traffic is challenging.} A natural question is whether colliding collectives can be avoided through application-level scheduling. Prior work~\cite{10.1145/3651890.3672249,285084,10.1145/3718958.3750499} synthesizes optimized schedules to reduce bandwidth contention within a single DC. While effective in controlled settings, these approaches face key limitations in our scenario. 
First, overlap between HAR and \ATOA is \textit{performance-driven}: modern frameworks~\cite{288705,10.1145/3458817.3476209} pipeline communication with computation to maximize utilization, so eliminating overlap directly reduces throughput. Second, existing schedulers assume fixed topology and predictable execution. These assumptions break in cross-DC deployments, where millisecond-scale RTTs, limited visibility, and variability in long-haul transfers introduce uncertainty that is difficult to capture statically. 
Thus, scheduling can reduce contention but cannot eliminate the residual interactions between concurrent cross-DC and intra-DC collectives at runtime.

\smartparagraph{In-switch deflection falls short under long bursts.} Prior work~\cite{abdous2023practical,zarifis2014dibs,gherghescu2024ve} avoids drops via in-switch deflection, targeting short microbursts in DCs, where packets are deflected to neighboring switches when buffers fill. These events last tens of microseconds, as senders react quickly to congestion. In contrast, in cross-DC transfers, senders observe congestion only at millisecond timescales, resulting in bursts that are orders of magnitude longer and repeatedly saturate shared links. Under these conditions, deflected packets make little forward progress: once rerouted, they are deprioritized and repeatedly deferred, circulating in the network and being \textit{unnecessarily deflected multiple times}. Rather than relieving congestion, deflection spreads it, increasing buffer occupancy and path lengths, and eventually causing drops when buffers fill~\cite{abdous2021burst}. Thus, while effective for microbursts, deflection breaks down under synchronized large-scale communication.

\smartparagraph{Desiderata.} Based on the above analysis, we seek a solution that satisfies the following requirements:
\begin{enumerate}[leftmargin=*,noitemsep,topsep=0pt,parsep=0pt,partopsep=0pt,label=D\arabic*]
    \item\label{desiderata1} \textit{Lossless recovery}: The system should eliminate packet drops, ensuring long-haul transmissions  avoid costly retransmissions over the cross-DC link.
    \item\label{desiderata2} \textit{Minimize deflection overhead}:  Deflected packets should not continuously circulate in the network. 
    \item\label{desiderata3} \textit{Practical deployability}: The solution should require no architectural changes 
    and should be deployable using current hardware, possibly with minimal changes.
\end{enumerate}

\smartparagraph{Opportunity: absorbing deflected packets avoids drops and unnecessary retransmissions.} We seek a design that satisfies the three desiderata. The key observation is that once a packet has traversed the long-haul link, it must not be dropped and should remain close to its destination (\ref{desiderata1}). This rules out loss-based approaches, which rely on costly retransmissions. Deflection can avoid drops, but uncontrolled rerouting spreads congestion, increases buffer pressure, and eventually causes drops elsewhere, violating desiderata~\ref{desiderata2}. Finally, we aim to preserve current deployments without requiring changes to NICs or transport behavior, ruling out approaches that depend on hardware modifications 
(\ref{desiderata3}).

The only viable approach is to combine deflection with controlled buffering: packets are redirected, held at a safe location, and released once congestion clears. This yields a simple design that augments the DC with external memory acting as an extension of switch buffers.

\noindent In summary, we tackle the following question: \begin{tcolorbox}[colback=gray!15,colframe=gray!15,
    left=1pt,right=1pt,top=1pt,bottom=1pt]
    \centering
    ``Can we prevent cross-DC packet drops under overlapping collectives, without creating new bottlenecks?''
\end{tcolorbox}

\section{Design}\label{sec:design}

We present \systemname, a system that eliminates packet drops after the long-haul link by turning burst absorption into a disaggregated buffer. Packets are deflected to \textit{spillway nodes}, which act as external buffers that hold traffic and later reinject it once the destination port has drained, avoiding costly retransmissions.
%

\smartparagraph{Design guarantees.} \systemname is designed to satisfy three key invariants: \first~packets that traverse the DCI are not dropped within the destination DC; \second~deflected packets do not continue circulating in the network but are instead held until it is safe to forward; and \third~congestion feedback to the sender is preserved despite a possible deflection.

\begin{figure}[b]
    \vspace{-.2in}
    \centering
    \includegraphics[width=.8\linewidth]{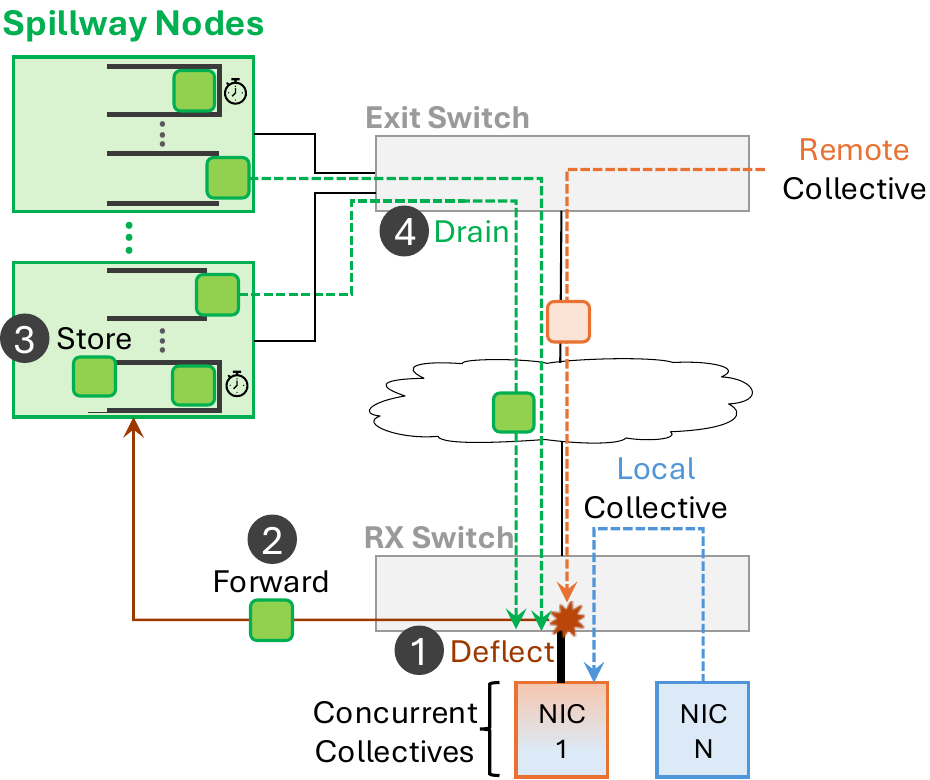}
    \caption{Overview of the \systemname architecture.}

    \label{fig:arch}
\end{figure}

\smartparagraph{Overview.} Fig.~\ref{fig:arch} illustrates an overview of the \systemname architecture. Consider a packet that arrives at a switch and needs to be forwarded toward a given egress port, but the corresponding buffer is congested due to competing traffic. The packet would normally be dropped. Instead, the switch \circleddark{1}~invokes \emph{deflect-on-drop}: it encapsulates the packet and redirects it toward a spillway node, load-balancing between multiple options. The encapsulated packet is \circleddark{2}~forwarded to the spillway node using standard DC routing. Upon arrival, the spillway decapsulates the packet and \circleddark{3}~stores it in a set of queues backed by a buffer. The spillway monitors each queue independently. Once no new packets have arrived for a configurable \textit{quiet interval}, the queue probes the path by transmitting the head-of-line packet. If the probe is not deflected back, and no new deflected packets arrive, the spillway \circleddark{4}~gradually releases the remaining buffered packets towards their destination.

\smartparagraph{Challenges.} \systemname deflects \textit{only} cross-DC HAR traffic: dropping a packet that has crossed the long-haul link would trigger a multi-millisecond retransmission that directly inflates iteration time, while local \ATOA sits on the microbatch critical path and must not be delayed by deflection (see Sec.~\ref{sec:background}). This asymmetric treatment is also the source of the core difficulty (Fig.~\ref{fig:arch}): the destination port must blend three traffic classes: high-priority local collectives, low-priority cross-DC HAR, and retransmissions from spillways. \systemname must ensure that draining buffered packets does not recreate the same incast conditions at the receiver.

This introduces three key challenges: \first~spillway nodes must be placed so that deflected packets can be offloaded quickly without traversing already congested links; \second~the system must decide when to drain buffered packets without direct visibility into the destination state: draining too aggressively across multiple spillways can recreate incast; and \third~deflected traffic must be distributed across spillways to avoid overloading individual nodes. 

We address these challenges by designing \systemname along three dimensions: \textit{where} to place spillway nodes, \textit{when} to drain buffered packets, and \textit{how} to distribute deflected traffic across spillways.

\subsection{\textit{Where} to Locate the Spillways}\label{ss:design-where}

A first important question is where to attach spillway nodes within the DC. As mentioned in Sec.~\ref{sec:motivations}, congestion often manifests at leaf switches, where cross-DC flows arriving from the DCI collide with local collective traffic at the same leaf that serves the destination GPUs. Spillway nodes must therefore be reachable from the leaf layer, but their placement must not stress the already-congested links.

Attaching spillway nodes at the leaf-GPU layer is the most intuitive option, but it is immediately ruled out: all downward-facing ports of a leaf switch are occupied by NICs connected to GPUs. Adding spillway nodes at this layer would require reducing the compute infrastructure, which is not a viable option.
Spine switches are one hop above the leafs and always reachable via normal routing. However, spine switches are also fully subscribed, \ie all their ports connect leaf switches downstream and exit switches upstream, leaving no spare capacity for spillway attachment. 

Exit switches stand apart. As discussed in Sec.~\ref{sec:motivations}, modern DC deployments rely on commodity exit switches to connect the internal fabric to the DCI. These switches operate under controlled oversubscription~\cite{meta_prometheus,gangidi2024rdma}: their aggregate DCI-facing bandwidth is intentionally lower than their intra-DC-facing bandwidth, leaving spare ports available. Thus, \systemname attaches spillway nodes to these spare ports. A packet deflected at a leaf is routed via the spine to an exit switch, and from there delivered to a locally-attached spillway in one additional hop. Critically, this path does not traverse any link already saturated by collective traffic: the spine-to-exit segment is dedicated to DCI-bound traffic, which is orthogonal to the east-west local traffic.

\subsection{Knowing \textit{When} to Drain a Spillway}\label{ss:design-when}

Once a packet is held at a spillway node, the spillway must decide when to retransmit it toward the original destination. Draining too aggressively can recreate incast at the destination, while draining too conservatively could inflate FCTs. The spillway has no direct visibility into the occupancy of the destination port, so it must infer this from indirect signals. 

A simple approach is to wait a fixed time $T$ before draining. However, the correct $T$ is workload-dependent and must match the duration of the competing collective, which varies across phases and scales. As a result, draining is decoupled from the actual network state, making it fragile under variability.
An alternative is to continuously drain at a rate adapted to arrivals: slowing under frequent deflections and speeding up otherwise. While reactive, this approach never fully pauses injection, and under heavy congestion continues to feed a saturated link, causing repeated deflections. This leads to persistent bouncing and wasted bandwidth.

\systemname instead leverages an already-available implicit signal: deflect-on-drop. When the spillway retransmits a packet, continued congestion causes it to be deflected back, while successful delivery results in no return. The spillway can thus infer the destination state purely from arrivals, without modifying switches or end-hosts. A natural design is to wait for a \textit{quiet interval} $\tau_{\mathrm{gap}}$, \ie a gap with no returning packets, and then drain the buffer in a burst. But, this approach is overly aggressive. Multiple spillways might observe the same gap and transmit simultaneously, so their combined bursts recreate the incast at the destination, triggering new deflections. 

To avoid this, \systemname uses a multi-step drain. After the quiet interval, the spillway's queue sends its head-of-line packet as a \textit{probe}. If the probe is deflected, it is re-buffered, the quiet interval resets, and the attempt is deferred. If the probe is not returned, the spillway sends a \textit{half-burst} as a conservative second test. If no packets are deflected, it escalates to a \textit{full burst}, restoring line-rate transmission. To further avoid synchronized drains, the quiet interval is randomized by adding a small jitter $\epsilon$ to $\tau_{\mathrm{gap}}$.


\subsection{\textit{How} to Balance across Spillways}\label{ss:design-how}

Distributing deflected traffic evenly across spillways is essential: a spillway that receives more traffic than it can buffer will start dropping packets, defeating the purpose of the system. Similarly, if too many flows are forwarded towards the same spillway, they will create an incast on the port, causing drops.  

The simplest approach assigns each spillway a unique unicast address and selects one via hashing the flow tuple, yielding a deterministic mapping with no switch state. However, this is prone to polarization~\cite{280752}: many flows may map to the same spillway while others remain underutilized. Under synchronized traffic, this concentrates deflections on a single spillway, creating an incast toward it. 
%
A better approach is to assign all spillways in a DC a shared anycast address and rely on per-packet spraying, broadly deployed in AI training DCs~\cite{spectrumx_whitepaper}, to spread deflections along least-congested paths, eliminating polarization without switch state. However, this conflicts with the drain mechanism in Sec.~\ref{ss:design-when}. Spraying can split a flow across multiple spillways, each observing only a partial arrival pattern, leading to incorrect quiet-interval and probe decisions. Worse, returned probes may be routed to a different spillway than the sender, breaking the feedback loop and disrupting the graduated drain.

\systemname combines both approaches. On the first deflection, the switch selects a spillway via anycast, achieving load-balanced placement. Upon retransmission, the spillway embeds its identifier in a packet header field (\eg IPv4 identification). If the packet is deflected again, the switch uses this identifier to route it back to the same spillway via unicast. This ensures that the entire drain loop for a flow, including probes and bursts, remains confined to a single spillway, preserving the feedback signal while keeping initial placement stateless.

Note that this design requires minimal programmability at the switch, \ie the ability to map a packet field to an IP address. While this can be easily implemented on programmable switches, it may not be supported on fixed-function ASICs. In such cases, the anycast-only design remains a viable alternative and achieves comparable performance, as shown in Sec.~\ref{ss:eval-sim}. However, without per-flow ``stickiness'', the probe mechanism becomes slightly less effective. Assuming per-packet spraying distributes the packets evenly among spillways, the probing mechanism still works, but now stochastically. If the spillways sense silence, they all would send the probe, and receive deflected packets again if there is a congestion. We evaluate all three techniques.

\subsection{Handling CC Feedback under Deflection}\label{ss:design-fastcnp}

\systemname can interfere with the standard CC feedback loop when packets experience congestion in the source DC before crossing the long-haul link. Assuming an ECN-based scheme such as DCQCN~\cite{dcqcn}, ECN-marked packets normally propagate to the destination and trigger CNP generation at the receiver, allowing the sender to reduce its rate. However, with \systemname, packets may be deflected in the destination DC before reaching the receiver, suppressing the corresponding CNP. As a result, the sender may not react in time to congestion in the source DC and continue injecting traffic at a high rate, breaking the isolation between local and remote traffic.

To address this issue, \systemname implements an additional mechanism at the source exit switch: upon receiving an ECN-marked packet, the switch generates a CNP and clears the ECN marking to avoid duplicate notifications. Rather than waiting for the packet to reach the receiver, the exit switch closes the feedback loop immediately by signaling the sender. Deflected traffic is treated separately. To avoid spurious congestion signals, packets redirected to spillways are assigned to a distinct traffic class on which ECN marking is disabled. This prevents deflection-induced buffering from interfering with the standard CC feedback loop, ensuring that ECN reflects only congestion experienced along the original data path.

While this mechanism has been explored in prior work~\cite{niu2025themis,11366742}, it serves a dual purpose in \systemname: it preserves timely rate adaptation to local congestion, and decouples CC from deflection, ensuring correct rate control even when packets are redirected to spillway nodes. 

\subsection{Analysis}

\begin{figure}[b!]
    \centering
    \includegraphics[width=\linewidth]{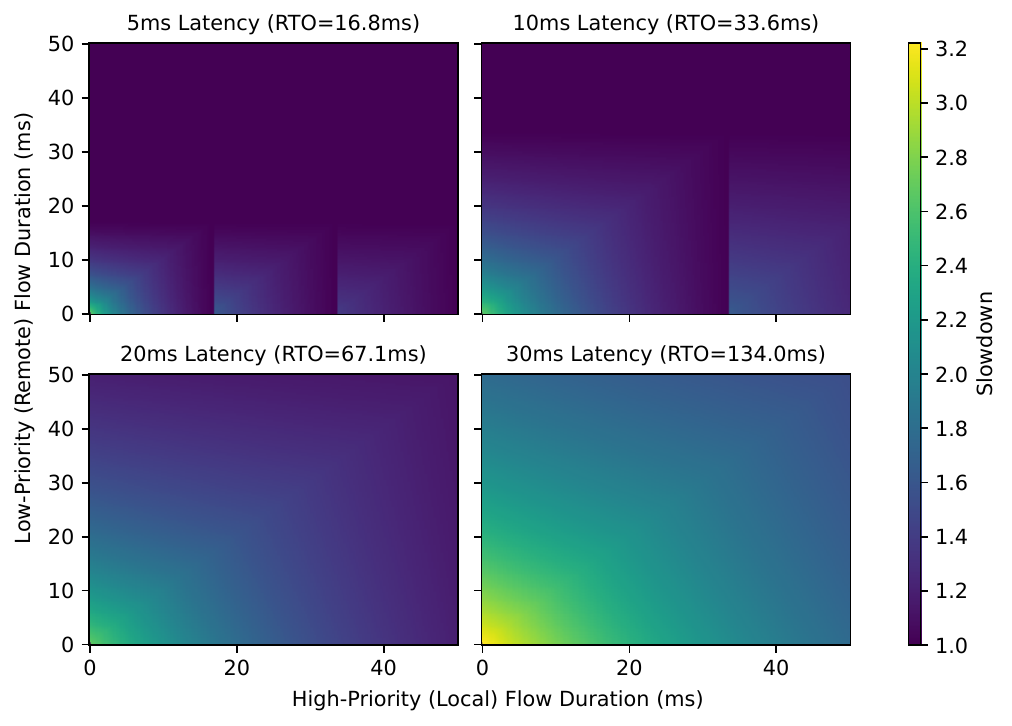}
    \vspace{-.25in}
    \caption{Cross-DC FCT slowdown vs.\ ideal under RTO-driven losses. Slowdown peaks for short flows and grows with link latency.}
    \label{fig:fct_model}
\end{figure}

We aim to understand the potential benefits of \systemname under different system and workload parameters. We build a simplified model that estimates the worst-case cross-DC FCT under RTO-driven recovery, where retransmissions incur an additional RTT of delay after the destination bandwidth becomes available. We compare this against an optimal solution with infinite buffers and perfect knowledge of when congestion ends.

We model two flows sharing a destination port: a cross-DC flow, and an intra-DC flow representing a local collective running in parallel. $L$ represents the one-way cross-DC delay, $T_r$ the transmission time of the cross-DC flow, and $T_a$ the transmission time of the local intra-DC collective. The local collective is prioritized and its FCT is unaffected by the cross-DC flow. If the destination link bandwidth is fully utilized by the local flow, arriving remote packets are dropped as soon as the switch buffer fills up. Retransmitted packets take one RTO to arrive which delays the FCT. The resulting remote FCT as a function of the above parameters, is as follows:
\[
\mathrm{FCT} \;=\;
\begin{cases}
T_r + T_a + \mathrm{RTT},
  & \mathrm{RTO} \le T_r \\[6pt]
T_a + \mathrm{RTO} + \mathrm{RTT},
  & \substack{\mathrm{RTO} > T_r,\\[2pt] (T_a \bmod \mathrm{RTO}) < T_r} \\[10pt]
\left\lceil \tfrac{T_a}{\mathrm{RTO}} \right\rceil \mathrm{RTO} + T_r + \mathrm{RTT},
  & \substack{\mathrm{RTO} > T_r,\\[2pt] (T_a \bmod \mathrm{RTO}) \ge T_r}
\end{cases}
\]
\[
\mathrm{RTO} = \alpha\mathrm{RTT}, \qquad \mathrm{RTT} = 2L,
\]
where $\alpha$ is some constant.

When $T_r \ge \mathrm{RTO}$, retransmissions incur no penalty: the remote flow is simply extended by the blocking duration $T_a$, matching the ideal overlap case. When $\mathrm{RTO} > T_r$, each loss incurs at least one full $\mathrm{RTO}$, leading to two sub-regions.
If $(T_a \bmod \mathrm{RTO}) < T_r$, the final retry partially overlaps with the local flow, and only the tail is dropped and retransmitted, yielding $\mathrm{FCT} = T_a + \mathrm{RTO} + \mathrm{RTT}$.
If $(T_a \bmod \mathrm{RTO}) \ge T_r$, the retry is dropped again, and completion occurs at the next retry, giving $\mathrm{FCT} = \lceil T_a/\mathrm{RTO} \rceil \mathrm{RTO} + T_r + \mathrm{RTT}$.
The trailing RTT in all cases accounts for the ACK arrival at the sender.

In both cases, FCT is driven by $\mathrm{RTO}$, which scales with the cross-DC latency. The ideal baseline is $\mathrm{FCT}_{\text{ideal}} = T_r + T_a + \mathrm{RTT}$, as its the earliest time the remote flow can complete, this time is used to compute the slowdown in Fig.~\ref{fig:fct_model}. We can see that for all link latencies, the baseline experiences the highest slowdown in the region where the remote and local flows are short, with the highest expected benefits from \systemname.

\subsection{Discussion}\label{ss:design-discussion}

\smartparagraph{Deployment model.} A key advantage of \systemname over alternative approaches is its limited impact on the existing stack. Deploying \systemname requires attaching spillway nodes to available ports on exit switches and configuring deflect-on-drop so that packets that would otherwise be dropped are redirected to spillways. These operations rely on standard switch capabilities and do not require changes to the training framework, collective scheduling, CC, or NIC/switch firmware.
The only functional assumption of \systemname is tolerance to OOO delivery, which already holds in deployments using per-packet adaptive routing, as is common in modern AI clusters~\cite{qian2024alibaba,gangidi2024rdma,ar_whitepaper,mcclure2026loadbalancingaitraining}.

\smartparagraph{Sizing the spillway buffer.} The spillway must buffer the aggregate cross-DC traffic that arrives while the destination port is monopolized by local traffic. The required capacity is therefore $B_{\mathrm{spillway}} \geq B_{\mathrm{agg}} \times T_{\mathrm{coll}}$, where $B_{\mathrm{agg}}$ is the aggregate arrival rate of the long-haul flows affected by the collision and $T_{\mathrm{coll}}$ is the duration of the competing local collective. As a concrete example, consider 16 long-haul flows, each sending at 400\,Gbps, and assume the DCI provides sufficient capacity so that these flows can be injected at line rate. If their aggregate traffic is blocked for $T_{\mathrm{coll}} = 5\,\mathrm{ms}$, the required buffer is $16 \times 400\,\mathrm{Gbps} \times 5\,\mathrm{ms} \times \frac{1}{8} = 4\,\mathrm{GB}$. This requirement remains well within the DRAM capacity of modern SmartNICs and commodity servers; \eg NVIDIA BlueField-3 provides 16\,GB of on-board memory~\cite{bf3}, making it straightforward to provision sufficient buffering. This capacity is scaled by the number of spillway nodes per exit switch, yielding an aggregate buffer pool on the order of hundreds of GBs. 

\smartparagraph{Setting the quiet interval.} \systemname exposes a single timing parameter: the quiet interval $\tau_{\mathrm{gap}}$.
The quiet interval defines how long the spillway waits after the last arrival before attempting to drain. It must exceed the RTT between the spillway and the destination leaf so that a deflected probe can return before the next attempt. Since the spillway is attached to an exit switch, this corresponds to the intra-DC RTT (typically 1-5\,$\mu$s), and a small multiple suffices in practice.
To avoid synchronized drains across spillways, the quiet interval is randomized by adding a small jitter $\epsilon$.

\smartparagraph{Ensuring progress under deflection.}
When a retransmitted packet encounters persistent congestion, it may be deflected again. This is intentional, as bounce-backs provide the feedback signal. However, it raises the risk of repeated circulation.
\systemname bounds this behavior with two mechanisms. First, unicast fallback ensures that re-deflected packets return to the \textit{same} spillway, avoiding cross-node cascades. Second, the drain logic limits retries: transmission occurs only after a quiet interval and a successful probe, while the deadline timer guarantees eventual progress. \systemname does not enforce strict packet ordering, as in OOO-tolerant deployments; supporting strict ordering is left for future work.

\section{Implementation}\label{sec:impl}

\smartparagraph{Simulator.} We build on the ASTRA-Sim framework~\cite{won2023astrasim2} using the ns-3 backend. Since ASTRA-Sim does not natively support multi-DC topologies and lossy RDMA, we extend the backend with both features. We further enhance the switch model to support two traffic classes: a lossless class with PFC and ECN, and a lossy class with ECN only, as well as per-packet spraying. Finally, we implement \systemname's logic by extending the switch model and introducing dedicated nodes that handle packet buffering and draining.

\smartparagraph{Hardware prototype.} We implemented \systemname on an experimental testbed consisting of a network switch performing the deflect-on-drop and a spillway node performing the buffering and controlled reinjection. 
%

We use an NVIDIA 51.2\,Tbps Spectrum-4 switch~\cite{spectrum4_ds} to perform the deflect-on-drop path: when a packet is tail-dropped at an egress queue, the switch deflects it toward the spillway node. The deflect-on-drop mechanism is realized using an existing packet mirroring capability that is traditionally used for telemetry and network analysis. We configure a mirror session on all of the switch's interfaces as the mirror sources and tie the mirror session to a tail-drop event, so any dropped packet on an egress queue is sent to the mirror session. The mirror session receives the packet untrimmed and encapsulates it with an L3 GRE header~\cite{rfc2784} targeting the spillway node's IP. In our prototype, we define a unicast IP address for the spillway node, but an anycast IP can be configured for delivery to multiple spillways, as we show in Sec.~\ref{ss:eval-testbed}.

The spillway node performs the buffering and controlled reinjection path. In our prototype we use an NVIDIA BlueField-3 single-port DPU~\cite{bf3}. The spillway runs as a DPDK application on the BF3 ARM processor.
%
Deflected packets first enter the RX HW pipeline, where the GRE header is decapsulated and Receive Side Scaling (RSS) steers each packet to one of four RX queues based on its original destination IP. This per-flow steering provides the isolation explained in  Sec.~\ref{sec:design}: traffic for different destinations is served by independent queues, so a deflection for one destination does not stall the drain of another. We configure four queues in our implementation, although the exact number is platform-dependent; for example, ConnectX-7-class NICs (as in the BF3) support up to 1024 RX queues per port.

Packets are received on per-queue RX queues, from which the application enqueues pointers into a software ring. The packet payloads reside in a shared \texttt{mbuf} pool allocated from the BF3's DRAM. This indirection decouples reception from transmission: packets may continue to arrive while the TX path is temporarily held back, and the ring provides buffering until transmission resumes.

Transmission follows the drain logic described in Sec.~\ref{ss:design-when}. Each queue tracks its own receive activity via a timeout; when no deflected packet has arrived for $\tau_{\mathrm{gap}} = 30\,\mu$s, the spillway issues a probe. If the probe is not deflected back, the spillway ramps to a half-rate burst, and then to full line rate if no further deflections are observed.



\section{Evaluation}\label{sec:eval}

In this section, we evaluate \systemname via ns-3 simulations (Sec.~\ref{ss:eval-sim}) and a real testbed (Sec.~\ref{ss:eval-testbed}).

\subsection{Simulation Evaluation}\label{ss:eval-sim}

\smartparagraph{Configuration.} We model two DCs connected via a DCI, each structured as a fat-tree topology. Each DC contains 32 GPUs. We assume eight GPUs per node~\cite{dgx_b200}, with each of the four nodes connected to a distinct leaf switch, resulting in four leaf switches per DC. The spine layer consists of eight switches, each fully connected to eight exit switches, matching the spine radix and ensuring no oversubscription at the DC boundary. All intra-DC links operate at 400\,Gbps with 1\,$\mu$s latency. At the DCI boundary, exit switches are paired across DCs (\ie the $i$-th exit of DC1 connects to the $i$-th exit of DC2). Each pair is interconnected via two 400\,Gbps links, providing 800\,Gbps of cross-DC capacity per exit switch. The one-way cross-DC latency is set to 5\,ms, corresponding to roughly 1000\,km of fiber distance, unless specified. When \systemname is enabled, four spillway servers (provisioned with 16\,GB of buffer capacity each) are connected to each exit switch. We set $\tau_{\mathrm{gap}} = 30\,\mu$s as in the DPDK prototype.

We model commodity switches with 64\,MB of shared buffer, prioritizing lossless over lossy queues and setting ECN thresholds per established guidelines~\cite{hpcc}. At the host, we use DCQCN~\cite{dcqcn} tuned for 400\,Gbps NICs, with lossless queue pairs similarly prioritized.



\smartparagraph{Workloads.} We use two workloads. The first models a realistic training iteration using a Chakra trace generated with MLSynth~\cite{10.1145/3748273.3749211}, extended to support HAR. The trace simulates a DeepSeek-V3-like sparse MoE model scaled to 24\,B parameters, with 0.6\,B active parameters per step. The model consists of 64 transformer layers (3 dense and 61 MoE) with hidden size 1024, where each MoE layer includes 128 experts with top-2 routing. The model is partitioned into 4 pipeline stages, with 4 data-parallel replicas and expert parallelism of 2. HAR is implemented using a bucketed approach, triggering an operation every two layers during the backward pass. The number of microbatches is 8.
The second workload is a microbenchmark derived from this trace, designed to isolate \systemname's behavior under controlled conditions. It captures the key communication patterns, \ie long-haul HAR transfers and intra-node \ATOA contention. Specifically, it consists of 16 long-haul lossy flows sent from DC1 to DC2, each of size 250\,MB (matching HAR chunk sizes that fill the BDP), along with concurrent intra-node lossless \ATOA operations of 4\,GB per node ($\approx$500\,MB per GPU). To emulate realistic variability in collective communication, we introduce controlled jitter in flow start times using a fixed random seed. 

\smartparagraph{Baseline.} We compare \systemname against an RDMA transport with out-of-order delivery support, where losses are recovered via RTO-based retransmissions. This reflects deployments in modern AI clusters~\cite{ar_whitepaper}.

\begin{figure}[t]
    \vspace{-.15in}
    \hfil
    \centering
    \subfloat[\centering Last microbatch duration of the straggler node.]{    	
        \includegraphics[width=.9\columnwidth]{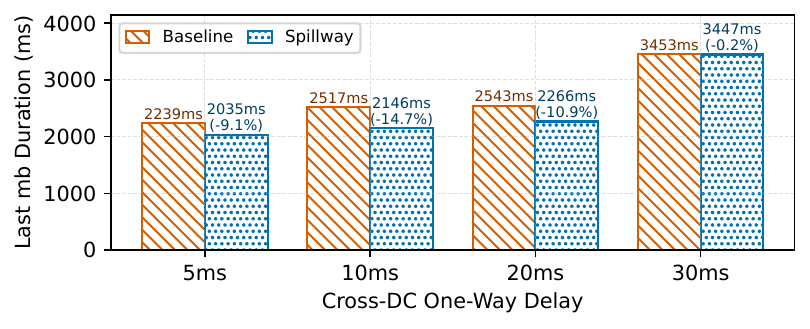}
    	\label{fig:workload_ub}
    }
	\hfil
    \centering
	\subfloat[\centering Iteration time improvement.]{
        \includegraphics[width=.9\columnwidth]{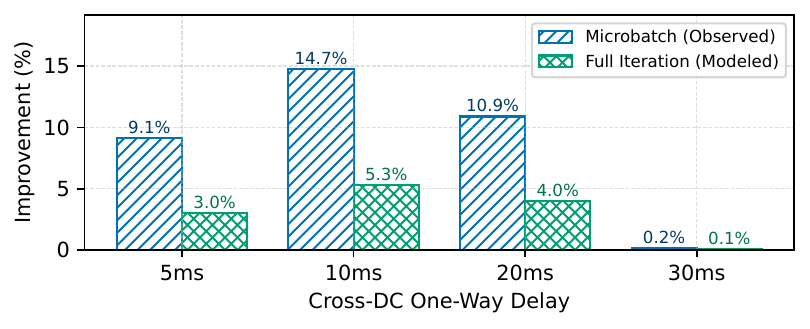}
        \label{fig:workload_iteration}
	}
    \caption{Impact of \systemname on training performance.}
	\label{fig:workload}
    \vspace{-.15in}
\end{figure}

\smartparagraph{\systemname reduces microbatch iteration time up to 14\%.}
We evaluate the end-to-end impact of \systemname on iteration time using the final microbatch of the Chakra trace. In this phase, gradients are computed and reduced both locally and across DCs via HAR, placing it directly on the critical path.
Fig.~\ref{fig:workload_ub} shows the microbatch duration of the straggler node for both the baseline and \systemname as we vary the one-way cross-DC delay from 5\,ms to 30\,ms ($\approx$1000-6000\,km), adjusting the retransmission timeout accordingly~\cite{ibta:spec}.
By buffering packets at the destination and avoiding retransmissions, \systemname reduces microbatch completion time up to 14\% (with 10\,ms delay). 

Note that the completion time for 10\,ms and 20\,ms is similar. This is not coincidental, but reflects a shift in the bottleneck. In the 20\,ms case, the cross-DC \AR is significantly delayed by the longer RTT, pushing subsequent phases later without introducing additional interference. In contrast, in the 10\,ms case, the AllGather phase overlaps with the \ATOA, introducing intra-DC contention. Because AllGather and \ATOA share the same lossless buffer, they compete at the destination leaf, triggering PFC and delaying the \ATOA. This delay propagates to the pipeline-parallel send stage and subsequent stages, offsetting the latency advantage. We provide a detailed analysis of this interaction in App.~\ref{app:har}.
At 30\,ms, our evaluation reveals a broader insight: local compute finishes before the cross-DC transfer, leaving no overlap for any buffering mechanism to exploit. We are the first to expose this effect. The evaluated trace retains healthy overlap up to 20\,ms; at higher RTTs, the parallelism should be re-tuned to better interleave compute with long-haul communication, which is beyond the scope of this paper and left as orthogonal and interesting future work.

We also extrapolate the microbatch gains to full iteration time by scaling the backward-pass duration as
$T_{\text{iteration}} = 1.5 \cdot t_{\text{bwd}}^{\text{stage}} \cdot (pp + mb - 1)$,
where the factor $1.5$ approximates the forward pass as shorter than the backward pass, consistent with prior observations that backward computation dominates training time~\cite{10.1145/3458817.3476209}. 
As shown in Fig.~\ref{fig:workload_iteration}, \systemname's benefits come from avoiding costly retransmissions and stabilizing cross-DC FCTs, yielding up to $\sim$5\,\% reduction. 

These percentages may appear modest but are substantial at scale: in multi-site deployments with $>$100\,K accelerators, even a 5\,\% reduction translates into \textit{thousands of GPU-hours saved} per iteration. Beyond raw compute savings, this directly improves cluster utilization and reduces energy consumption and operational cost. At hyperscale, where training efficiency is increasingly constrained by power budgets and infrastructure limits rather than GPU availability, such gains translate into meaningful reductions in total energy footprint~\cite{huber2025energyconsumptionparallelneural}.


\smartparagraph{Microbenchmark: anycast spillway selection avoids drops.} 
We evaluate the three spillway selection strategies (Sec.~\ref{ss:design-how}): \first~\textit{DC-Anycast}, where all spillway servers within a DC share a single anycast address; \second~\textit{SW-Anycast}, where each group of spillway servers (\ie the four connected to the same exit switch) shares a distinct anycast address; and \third~\textit{Unicast}, where each spillway server is assigned a unique unicast address. For each strategy, we further consider two variants: \textit{Sticky} (first anycast, then unicast) and \textit{Stateless} (anycast only).

\begin{figure}[t]
    \centering
    \includegraphics[width=.85\linewidth]{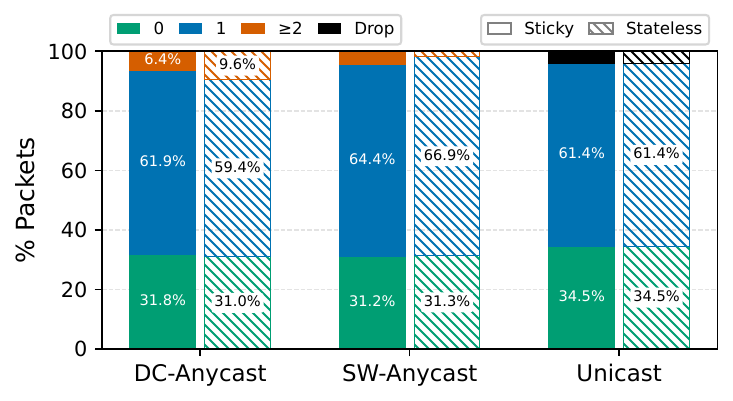}
    \caption{Deflection distribution with selection strategies.}
    \vspace{-.15in}
    \label{fig:spmb_histogram}
    \vspace{-.1in}
    
\end{figure}

Fig.~\ref{fig:spmb_histogram} reports the distribution of deflections across the different strategies. A first key observation is that pure unicast approaches are ineffective, as they lead to packet drops, highlighting the need for anycast to balance load. Focusing on the anycast approaches, both achieve similar behavior for packets experiencing a single deflection (roughly 60\%). Differences emerge in the tail of the distribution, where SW-Anycast results in fewer packets experiencing multiple deflections. This can be attributed to the smaller selection domain: choosing among a subset of four spillways increases the likelihood that packets are consistently directed to the same destination, as opposed to selecting from a larger pool. Comparing Sticky and Stateless variants, we observe no significant differences between the two, indicating that both approaches are viable depending on the switch hardware capabilities.

\begin{figure}[b!]
    \vspace{-.25in}
    \hfil
    \centering
    \subfloat[\centering DC-Anycast.]{    	
        \includegraphics[width=.9\columnwidth]{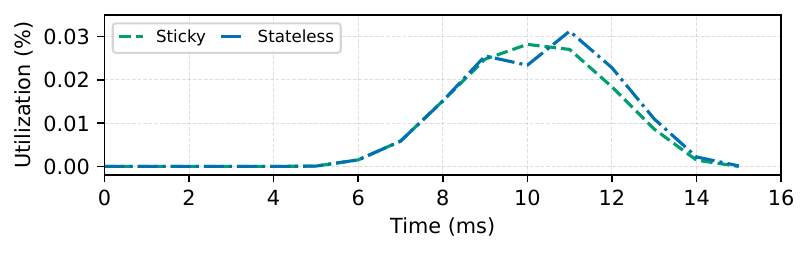}
    	\label{fig:spmb_util_dca}
    }
	\hfil
    \centering
	\subfloat[\centering SW-Anycast.]{
        \includegraphics[width=.9\columnwidth]{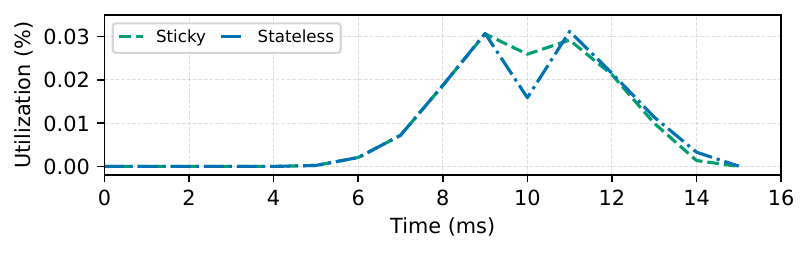}
        \label{fig:spmb_util_swa}
	}
    \caption{Spillway buffer utilization over time.}
	\label{fig:spmb_util}
\end{figure}

Fig.~\ref{fig:spmb_util} reports spillway buffer utilization for the two anycast strategies, normalized to the aggregate capacity (\ie 512\,GB). Utilization remains low in all cases. The Stateless variant drains more aggressively, occasionally triggering small incasts that cause limited re-deflections; however, this effect is bounded (at most four deflections) and has negligible impact.

Overall, anycast is essential to avoid drops, while both Sticky and Stateless designs are viable.

\begin{figure}[t]
    
    \centering
    \includegraphics[width=.95\linewidth]{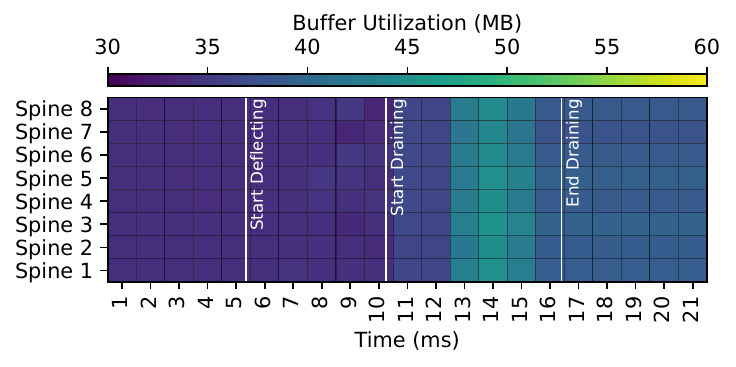}
    \vspace{-.15in}
    \caption{Spine buffer utilization under extreme congestion.}
    \label{fig:xpsw_buffer}
    \vspace{-.2in}
    
\end{figure}

\smartparagraph{Microbenchmark: spillway deflection remains robust under extreme spine congestion.} A concern is that deflected packets may be dropped before reaching spillway servers if the spine is congested, or that deflection itself adds excessive load. We show that \systemname avoids both issues. We stress the spine by injecting continuous 400\,Gbps UDP traffic from each GPU in DC2 toward unused servers\footnote{The topology has 32 GPUs per DC, only 16 participate in the \ATOA.}, on top of the base workload. These flows are uncontrolled and can be dropped, serving only to saturate the spine layer.
Fig.~\ref{fig:xpsw_buffer} shows that spine buffers remain around 30\,MB during the collision phase. When deflection starts ($\sim$5\,ms), buffer occupancy increases only slightly. During draining ($\sim$10\,ms), it peaks at 45-50\,MB due to reinjection, then quickly returns to low levels.
\begin{figure}[b]
    \vspace{-.2in}
    \centering
    \includegraphics[width=.95\linewidth]{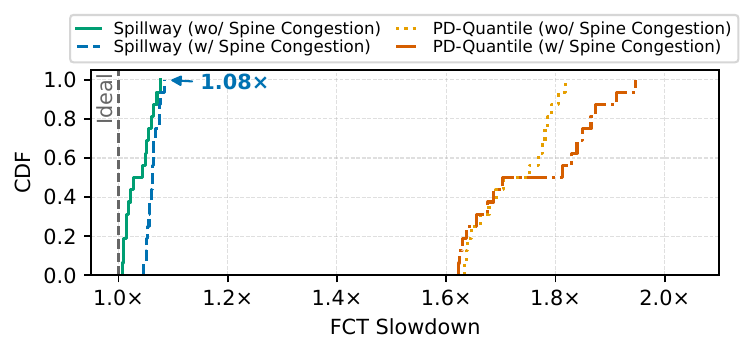}
    \vspace{-.1in}
    \caption{FCT slowdown under spine congestion.}
    \label{fig:xpsw_cdf}
\end{figure}
We further quantify the FCT slowdown under this stress scenario to assess its performance impact. We compare \systemname with a representative of in-switch deflection, \ie preemptive deflection~\cite{abdous2023practical}, with parameters configured as in the original work. Fig.~\ref{fig:xpsw_cdf} reports the FCT slowdown of both systems, with and without spine congestion, normalized to the ideal FCT of the 16 flows. \systemname maintains minimal slowdown across both configurations, with a maximum of 1.08$\times$ even under spine congestion. In contrast, preemptive deflection degrades significantly, reaching $\sim$2$\times$.

We emphasize that this scenario represents an extreme and unlikely operating condition, and is primarily used to evaluate the robustness of \systemname under severe congestion.

\smartparagraph{Microbenchmark: fast CNP is essential to preserve CC feedback under deflection.} As described in Sec.~\ref{ss:design-fastcnp}, \systemname can disrupt the standard CC feedback loop when packets experience congestion in the source DC. 
In this microbenchmark, we evaluate the benefits of fast CNP generation at the source exit switches. To induce local congestion, we reduce the DCI link bandwidth from 800\,Gbps to 400\,Gbps so that when multiple flows are routed through the same exit, they experience contention and must reduce their sending rate to share the available capacity. 

\begin{figure}[b!]
    \vspace{-.3in}

    \hfil
    \centering
    \subfloat[\centering CNPs from exit switches.]{    	
        \includegraphics[width=.9\columnwidth]{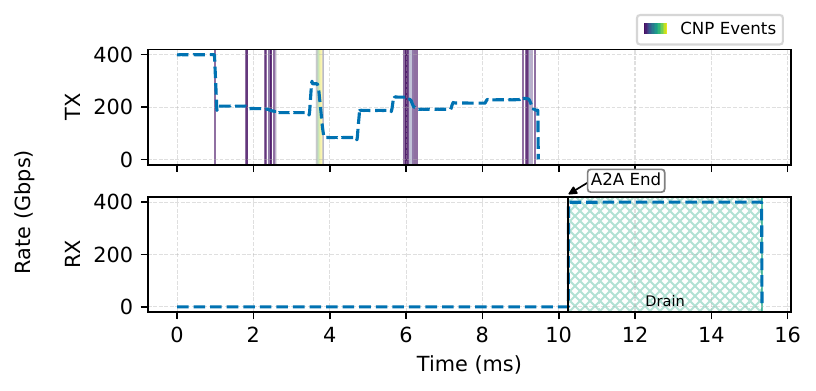}
    	\label{fig:concurrent_fastcnp}
    }
	\hfil
    \centering
	\subfloat[\centering CNPs from the receiver.]{
        \includegraphics[width=.9\columnwidth]{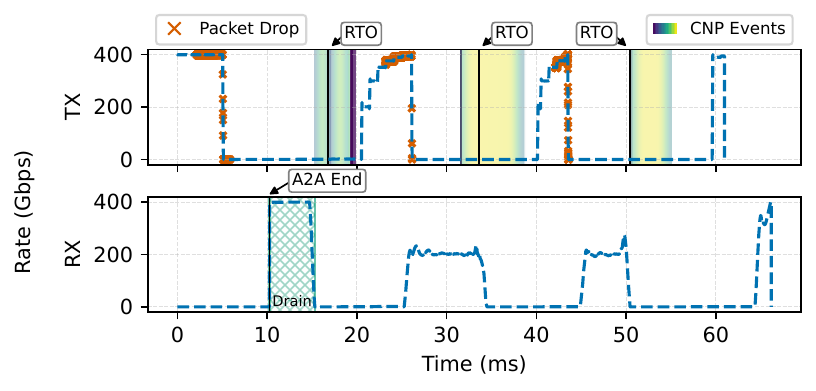}
        \label{fig:concurrent_nofastcnp}
	}
    \caption{Impact of fast CNP feedback on flow behavior.}
	\label{fig:concurrent_cnp}
\end{figure}

In Fig.~\ref{fig:concurrent_cnp}, we report the rate in Gbps (y-axis) over the experiment time (x-axis) for one of the 16 remote flows, as observed at both the transmission (TX) and receiving (RX) sides. We evaluate the system both with (Fig.~\ref{fig:concurrent_fastcnp}) and without (Fig.~\ref{fig:concurrent_nofastcnp}) fast CNP feedback from the exit. The heatmaps represent the number of CNPs received by the sender in 1\,ms intervals. As observed, enabling fast CNPs allows the flow rate to adjust promptly, thanks to the intra-DC feedback loop operating on the order of microseconds. The rate quickly converges to approximately 200\,Gbps, indicating that the flow shares a DCI port with another flow. On the receiving side, packets are buffered until the lossless \ATOA completes, after which they are drained from the spillways (x-hatched green box), allowing the flow to finish in about 20\,ms, close to the ideal FCT. Conversely, without fast CNPs, the flow does not react promptly to congestion, resulting in multiple packet drops at the exit switches ($\times$ orange marks). Although transmission completes, packets remain buffered for an additional 10\,ms before they can be retrieved from the spillway. At that point, a burst of CNPs is generated, forcing the flow rate down to the minimum DCQCN value. The rate then increases again, but without effective regulation, further packet drops occur at the DCI boundary. While packets that reach the destination eventually trigger CNP feedback, this arrives too late to prevent losses. This cycle repeats until all packets are delivered, inflating the FCT to approximately 70\,ms. Overall, fast CNP feedback is critical to maintain effective CC under deflection, preventing loss amplification.

\vspace{-.1in}
\subsection{Testbed Evaluation}\label{ss:eval-testbed}

\smartparagraph{Configuration.} Our testbed consists of an NVIDIA Spectrum-4 switch and a spillway node implemented on an NVIDIA BlueField-3 DPU (8 cores for 100\,Gbps), along with dual-port and single-port ConnectX-6 Dx NICs acting as sender and receiver, respectively.
Links operate at 100\,Gbps. To isolate the spillway's behavior from CC dynamics, we disable CC on all endpoints, sustaining line-rate traffic at the destination. This emulates cross-DC senders whose multi-millisecond feedback loop has not yet reacted, so any successful delivery is attributable directly to the spillway.
To emulate intra-\slash cross-DC traffic, we configure three traffic classes. On the dual-port NIC, one port marks packets with DSCP mapped to priority~3 and the other to priority~1. The switch is configured accordingly, mapping priority~3 to a lossless class with PFC (as in RoCEv2) and priority~1 to a lossy class representing cross-DC traffic. Packets drained from the spillway are re-marked to priority~2, isolating them from original traffic.
Experiments are repeated $10$ times.

\begin{figure}[t!]
    \centering
    \includegraphics[width=.9\linewidth]{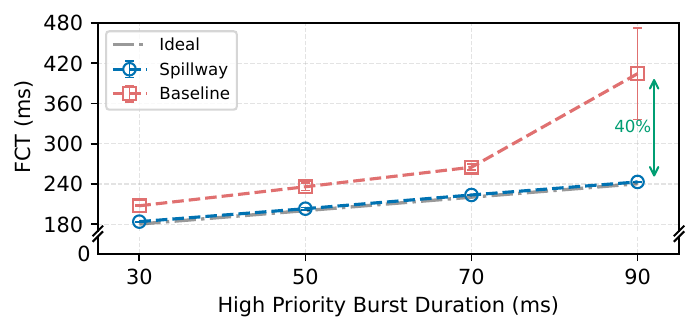}
    \vspace{-.1in}
    \caption{FCT of the lossy flow vs.\ high-priority burst.}
    \label{fig:testbed_baseline}
    \vspace{-.15in}
\end{figure}

\smartparagraph{\systemname avoids drops and approaches ideal FCT.} This experiment validates the core functionality of \systemname: packets are deflected under congestion, buffered at the spillway, and reinjected once the burst subsides. We generate a single low-priority RDMA flow, while the high-priority port injects periodic bursts with 270\,ms gaps. Due to strict priority, the low-priority flow is blocked and its packets are deflected.
Since the testbed lacks high-latency links, retransmissions would occur quickly and mask the spillway's behavior. To isolate the deflect-buffer-drain path, we disable retransmissions on the lossy QP. As a result, any packet dropped at the switch is permanently lost, and successful completion of the flow directly reflects recovery via the spillway.
Fig.~\ref{fig:testbed_baseline} shows the average FCT of the low-priority flow (y-axis) as a function of the high-priority burst duration (x-axis). We compare \systemname against a drop-based baseline with a 33\,ms RTO (simulating a long-haul of 10\,ms). The baseline (orange) is highly sensitive to burst duration, reaching an average FCT of 404\,ms with 90\,ms bursts. In contrast, \systemname (blue) recovers all deflected packets during the burst, achieving an FCT of 243\,ms under the same conditions, close to the ideal and a 40\,\% reduction over the baseline.

\begin{figure}[h!]
    \vspace{-.1in}
    \centering
    \includegraphics[width=.9\linewidth]{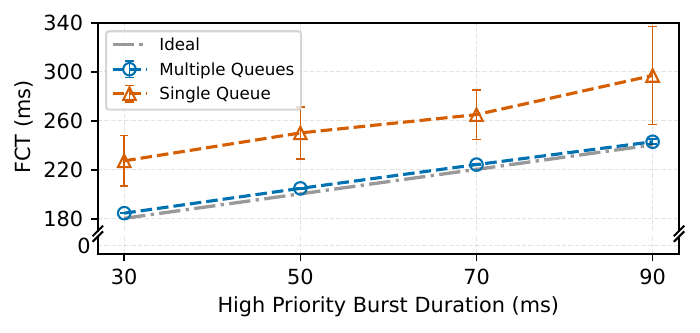}
    \vspace{-.1in}
    \caption{FCT of the lossy flow vs.\ high-priority burst with single and multiple queues in the spillway.}
    \label{fig:testbed_mq_vs_sq}
    \vspace{-.1in}
\end{figure}

\smartparagraph{Multi-queue isolation prevents cross-flow interference in the spillways.} In this experiment, we validate the per-flow queue design implemented via RSS on the destination IP. We extend the first setup by adding a NIC that generates congestion at a second switch port. This NIC sends constant traffic at half line rate to a second destination, while the dual-port NIC injects 50\,ms bursts every 120\,ms toward the same port. During each burst, the combined rate saturates the link, causing packets to be deflected to the spillway. As a result, the spillway handles deflected traffic from two independent flows: the original ``remote'' flow and the bursty interfering flow.
We compare two spillway configurations: a \emph{single-queue} variant where all flows share one RX/TX queue pair, and a \emph{multi-queue} variant where RSS assigns each flow its own queue.
Fig.~\ref{fig:testbed_mq_vs_sq} shows the average FCT (y-axis) as a function of the high-priority burst duration (x-axis). In the single-queue design (orange), deflected packets from the interfering flow share the same queue, so each deflection resets the quiet interval and delays probing, even if the flow under test is no longer congested. As a result, its completion time depends on the timing of the interfering bursts, leading to high variance. In contrast, the multi-queue design (blue) isolates flows across queues: the interfering traffic drains independently, and the flow under test matches the single-flow baseline.

\smartparagraph{Anycast distributes deflections evenly across spillways.} The design in Sec.~\ref{ss:design-how} uses anycast addressing with per-packet spraying to distribute deflected traffic across spillways without per-flow state. We validate this on our switch by advertising the same anycast route on four ports and injecting traffic toward it. With spraying enabled, 100\,K packets are evenly distributed, with \mbox{per-port} counts within 1\,\% of the mean, confirming balanced placement.
\vspace{-.1in}
\section{Related work}\label{sec:related}


\smartparagraph{Cross-DC LLM training.} Industry experience highlights the difficulty of managing congestion in multi-datacenter LLM training. 
Meta~\cite{gangidi2024rdma} reports disabling DCQCN after unsuccessful tuning, while Alibaba~\cite{qian2024alibaba} designed a custom CC scheme for this setting. 
These results indicate that tuning existing mechanisms is insufficient in practice. 
Our work builds on this observation by identifying the underlying structural causes and addressing them without modifying existing transport or collective stacks.

\smartparagraph{Cross-DC transfer mechanisms.} Prior work extends RDMA across DCs and mitigates long-haul congestion. Annulus~\cite{saeed2020annulus} and BiCC~\cite{wan2024bicc} address RTT-induced unfairness, while Zhou et al.~\cite{zhou2025mitigating} accelerate convergence under cross-DC incast. Other efforts (\eg IRN~\cite{mittal2018revisiting}) improve RDMA in lossy or high-RTT settings. These approaches assume independent flows and rely on reactive control. In contrast, cross-DC LLM training induces synchronized bursts that deterministically create congestion, where feedback or extra bandwidth is insufficient. \systemname instead targets this structural coupling and avoids loss amplification.

\smartparagraph{Exit-based CCs.} Several works regulate traffic using signals from exit switches~\cite{niu2025themis,wan2024bicc,11366742,11366779}, exploiting the fact that congestion often appears at DC boundaries. These designs maintain per-flow state at a single exit to control injection rate or cap in-flight data. However, they assume each flow traverses a single exit. This breaks under packet spraying~\cite{ar_whitepaper,gangidi2024rdma,qian2024alibaba,mcclure2026loadbalancingaitraining}, where packets of the same flow may traverse multiple exits, preventing any switch from maintaining a consistent view. As a result, flow-level control cannot be enforced, making these approaches inapplicable in our setting.

\smartparagraph{Packet deflection in DCs.} Prior work explores packet deflection to mitigate transient congestion within DC fabrics. DIBS~\cite{zarifis2014dibs} proposes just-in-time deflection to absorb microbursts via alternate output ports. Vertigo~\cite{abdous2021burst} and preemptive deflection~\cite{abdous2023practical} further refine this design space with selective deflection and controlled reordering to maintain performance under realistic Clos topologies. 
These systems operate within a single fabric and rely on alternative in-fabric capacity to absorb localized congestion. In contrast, cross-DC training induces synchronized bursts that repeatedly saturate cross-DC boundary links and couple multiple fabrics, a regime not addressed by prior in-fabric deflection mechanisms.

\vspace{-.15in}
\section{Conclusion}

We presented \systemname, a simple design that eliminates drops by turning burst absorption into disaggregated buffering. By operating entirely within the DC, \systemname enables microsecond-scale control without changes to end hosts or transports. Simulations show up to 14\,\% iteration time reduction, while the hardware prototype achieves up to 40\,\% FCT reduction. Our results highlight disaggregated buffering as a promising approach to decouple buffer capacity from fixed switch resources, which may be applied to a wide range of DC workloads.
Finally, we are the first to expose the challenges of overlapping communication and computation in multi-DC training under high RTTs. Addressing this interaction is an important direction for future work.

\bibliographystyle{plain}
\bibliography{references}
\appendix

\section{Cross-DC Latency Mitigating Intra-DC HAR Contention}\label{app:har}

Hierarchical AllReduce (HAR) proceeds in three phases: an intra-DC Reduce-Scatter, a cross-DC reduction on the resulting shards, and an intra-DC AllGather. The Reduce-Scatter operates across intra-DC data-parallel replicas, partitioning gradients to reduce cross-DC traffic. The cross-DC phase aggregates these shards across sites, and the AllGather redistributes the final result to all replicas. This process introduces additional intra-DC traffic which, as discussed in Sec.~\ref{sec:background}, can interfere with collectives on the critical path.

\begin{figure}[h!]
    \centering
    \includegraphics[width=\columnwidth]{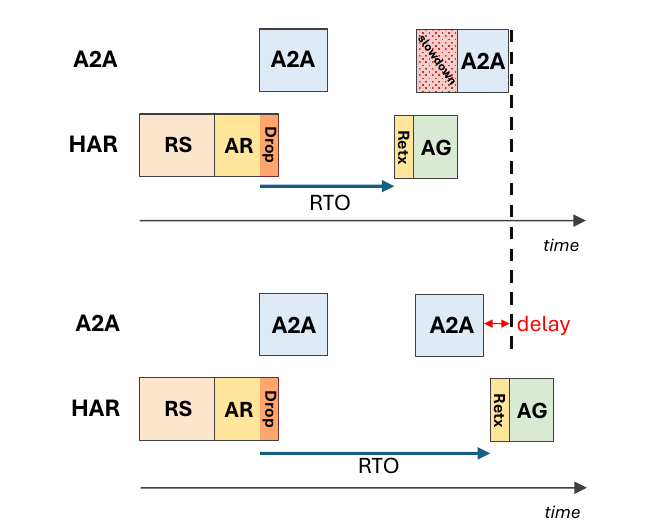}
    \caption{\ATOA collective is delayed due to overlap with HAR collective caused by short retransmission time.}
    \label{fig:appendix}
\end{figure}

In our workload simulations, we observe a non-intuitive interaction between cross-DC latency and iteration time. During the final microbatch, HAR overlaps with the backward pass, and its intra-DC phases can contend with \ATOA collectives on the critical path. Counterintuitively, increasing cross-DC latency can reduce this contention: the cross-DC reduction stretches over time, separating the two intra-DC HAR phases and reducing their overlap with critical \ATOA traffic. As shown in Fig.~\ref{fig:appendix}, iteration time can therefore remain nearly constant as cross-DC latency increases. While \systemname introduces some additional contention by filling idle communication gaps, the benefit of avoiding costly retransmissions dominates overall.

\end{document}